\documentclass{article}
\usepackage{graphicx}
\usepackage{amssymb}
\usepackage[cmtip,arrow]{xy}
\usepackage{pb-diagram,pb-xy}
\newcommand{\bPf}{\par\vspace*{-4pt}\indent{\sc Proof.}\enskip}
\newcommand{\ePf}{\medskip}
\newtheorem{corollary}{Corollary}
\newtheorem{remark}{Remark}
\newtheorem{lemma}{Lemma}
\newtheorem{proposition}{Proposition}
\newtheorem{theorem}{Theorem}
\def\QED{\hskip0.1em\hfill\null\ \null\nobreak\hfill\kern3pt\vbox{\hrule\hbox
   {\vrule\kern1pt\vbox{\kern1.7pt\hbox{$\scriptscriptstyle{QED}$}
    \kern0.2pt}\kern1pt\vrule}\hrule}}

\def\END{\hskip0.1em\hfill\null\ \null\nobreak\hfill\kern3pt\vbox{\hrule\hbox
   {\vrule\kern1pt\vbox{\kern1.7pt\hbox{$\,\,\,\vspace{5pt}$}
    \kern0.2pt}\kern1pt\vrule}\hrule}}
\newtheorem{definition}{Definition}
\newtheorem{example}{Example}
\newcommand{\ben}{\begin{enumerate}}
\newcommand{\een}{\end{enumerate}}
\newcommand{\bEq}{\begin{eqnarray}}
\newcommand{\eEq}{\end{eqnarray}}
\newcommand{\beq}{\begin{eqnarray*}}
\newcommand{\eeq}{\end{eqnarray*}}
\newcommand{\bDf}{\begin{definition}\em}
\newcommand{\eDf}{\end{definition}}
\newcommand{\bLm}{\begin{lemma}}
\newcommand{\eLm}{\end{lemma}}
\newcommand{\bPr}{\begin{proposition}}
\newcommand{\ePr}{\end{proposition}}
\newcommand{\bTh}{\begin{theorem}}
\newcommand{\eTh}{\end{theorem}}
\newcommand{\bCr}{\begin{corollary}}
\newcommand{\eCr}{\end{corollary}}
\newcommand{\bRm}{\begin{remark}\em}
\newcommand{\eRm}{\end{remark}}
\newcommand{\bEx}{\begin{example}\em}
\newcommand{\eEx}{\end{example}}

\newcommand{\ie}{{\em i.e$.$} }
\newcommand{\eg}{{\em e.g$.$} }
\newcommand{\R}{I\!\!R}

\newcommand{\cE}{\mathcal{E}}

\newcommand{\cL}{\mathcal{L}}

\newcommand{\bG}{{\bf G}}

\newcommand{\bK}{{\bf K}}

\newcommand{\bP}{{\bf P}}

\newcommand{\bX}{{\bf X}}
\newcommand{\bY}{{\bf Y}}

\newcommand{\alp}{\alpha}
\newcommand{\bet}{\beta}

\newcommand{\del}{\delta}
\newcommand{\eps}{\epsilon}

\newcommand{\lam}{\lambda}
\newcommand{\sig}{\sigma}

\newcommand{\For}{{\Lambda}}

\newcommand{\Var}{{\mathcal{V}}}
\newcommand{\Thd}{{\Theta}}

\title{\bf{ Cohomological obstructions in locally variational field theories}}

\author{Mauro Francaviglia\footnote{Mauro Francaviglia suddenly passed away in June 2013; this paper is the outcome of our recent joint research work and we dedicate it to his memory.},
Marcella Palese and Ekkehart Winterroth
\\
\footnotesize{Department of Mathematics, University of Torino, via C. A.lberto 10, I-10124 Torino , Italy}
\\
\footnotesize{e-mails: marcella.palese@unito.it, ekkehart.winterroth@unito.it}}

\date{}

\begin{document}
\maketitle

\begin{abstract}
We study cohomological obstructions to
the existence of global conserved quantities. In
particular, we show that, if a given local variational problem is supposed to admit global solutions, certain cohomology classes cannot appear as obstructions. Vice versa, we obtain a
new type of cohomological obstruction to the existence of global
solutions for a variational problem.
\medskip


\noindent {\em keywords}: jet, gauge-natural bundle, conserved quantities, cohomology.

\end{abstract}

\section{Introduction}

We are interested in studying corservation laws associated with invariance properties of {\em global} field equations which are {\em locally variational}, this means that locally they can be derived variationally from a Lagrangian, which is then only defined on an open set of the relevant field space;  this is what we call a {\em local presentation} of a given variational problem.
The requirement of {\em globality} for field equations, of course, determine the relationship between a Local Lagrangian defined on one open set and another local Lagrangian defined on another open set providing the same global field equations: 
we require two local presentations to differ by a variationally closed form, \ie locally by a variationally trivial Lagrangian (a local divergence).
Noether currents for such local presentations and corresponding {\em conserved currents associated with each local presentation}  have been characterized 
\cite{FePaWi10}; although {\em  they are associated with invariance of global field equations}, such conserved currents are, in principle, {\em local currents}.

In this paper we show that there exists cohomological obstructions for such local currents be globalized
and  we shall see how such obstructions are also related with the {\em existence of global solutions} for a given global field equation.
To investigate when local variational objects can be globalized, we use a geometric formulation of the calculus of variations as a subsequence of the de Rham sequence of differential forms on finite order prologations of fibered manifolds.

We assume the $r$-th order prolongation of a fibered manifold $\pi : \bY \to \bX$, with $\dim \bX = n$ and $\dim \bY = n+m$, to be the configuration space; this means that {\em fields are assumed to be (local) sections} of $\pi^r : J_r \bY \to \bX$.
Due to the affine bundle structure of  
$\pi^{r+1}_{r} : J_{r+1} \bY \to J_{r} \bY$, we have a natural splitting
$
J_r\bY\times_{J_{r-1}\bY}T^*J_{r-1}\bY =
J_r\bY\times_{J_{r-1}\bY}(T^*\bX \oplus V^*J_{r-1}\bY)$, which induces natural  { \em spittings} in  horizontal and vertical parts of vector fields, forms and of the exterior differential on $J_r\bY$.
Starting from this splitting one can define sheaves of contact forms $\Thd^{*}_{r}$ defined by the kernel of the horizontalization; the sheaves $\Thd^{*}_{r}$ form an exact subsequence of the de Rham sequence on $J_r\bY$ and one can define the quotient sequence
\beq
 0\to \R_{\bY} \to \dots
\to^{\cE_{n-1}} \For^{n}_r/\Thd^{n}_r \to ^{\cE_{n}} \For^{n+1}_r/\Thd^{n+1}_r  \to^{\cE_{n+1}}
\For^{n+2}_r/\Thd^{n+2}_r
\to^{\cE_{n+2}} \dots\to 0 \,
\eeq
the $r$--th order {\em variational sequence} on $\bY\to\bX$ which is an acyclic resolution of the constant sheaf $\R_{\bY}$; see \cite{Kru90}.

The quotient sheaves in the variational sequence can be represented as sheaves $\Var^{k}_{r}$ of $k$-forms on jet spaces of higher order, see \eg \cite{KrMu05,Kru90,KrSe05}.
Lagrangians are sheaf sections  $\lam\in(\Var^{n}_{r})_{\bY}$, while $\cE_n$ is called the Euler-Lagrange morphism. The Euler-Lagrange equations  are then $\cE_{n}(\lam)\circ j_{2r+1}\sig =0$ for (local) sections $\sig: \bX \to \bY$.
Sections $\eta\in(\Var^{n+1}_{r})_{\bY}$ are called {\em source forms} or  also {\em dynamical forms}, 
$\cE_{n+1}$ is called the Helmholtz morphism. The  equations $\eta \circ  j_{2r+1}\sig = 0$ are locally variational if and only if Helmholtz conditions $\cE_{n+1}(\eta)=0$ hold true.

\subsection{Inverse and local variational problems}
Since the variational sequence is exact as a sheaf sequence, it is possible to determine cohomological obstructions to global inverse problems: \ie given a closed section of a quotient sheaf of the variational sequence, when this section is also globally exact. 
In fact, the cohomology of the complex of global sections $H^{*}_{VS}(\bY)$ is naturally isomorphic to both  the \v Cech cohomology and the de Rham cohomology $H^{*}_{dR}(\bY)$ \cite{Kru90}.
Thus if the cohomology of $\bY$ is trivial, each local inverse problem is also global;
we are here interested in the case of a nontrivial cohomology of $\bY$.
Let $\bK^{p}_{r}\doteq \textstyle{Ker}\,\cE_{p}$; we have a natural short exact sequence of sheaves
which  gives rise in a standard way \cite{Bre67} to a long exact
sequence in \v Cech cohomology,
where the {\em connecting homomorphism} $\delta_{p} = i^{-1}\circ\mathfrak{d}\circ\mathcal{E}_{p}^{-1}$ is the mapping of cohomologies in the corresponding diagram of cochain complexes (here $\mathfrak{d}$ is the usual {\em coboundary operator}), see \eg \cite{BFFP03,FePaWi10,PaWiGa12}. 
Every $\eta\in(\cE_{p}(\Var^{p}_{r}))_{\bY}$ (\ie locally variational) defines a cohomology class 
$\del_{p} (\eta) \in
H^{1}(\bY, \bK^{p}_{r}) $ $\simeq$ $ H^{p+1}_{VS}(\bY)
$ $\simeq$ $ H^{p+1}_{dR}(\bY)$ (in particular, $\cE_{n}(\Var^{n}_{r})$ is the sheave of Euler--Lagrange morphisms and $\eta\in(\cE_{n}(\Var^{n}_{r}))_{\bY}$ if and only if $\cE_{n+1}(\eta)=0$, which are Helmholtz conditions).
In the following we shall use the notation  $\del (\eta)\equiv\del_{n} (\eta)$; we furthermore point out that
every $\mu\in(d_H(\Var^{n-1}_{r}))_{\bY}$ (\ie locally variationally trivial) defines a cohomology class 
$\del' (\mu) \equiv  \del_{n-1} (\mu)\in H^{1}(\bY,  \bK^{n-1}_{r}) $ $\simeq$ $ H^{n}_{VS}(\bY)
$ $\simeq$ $ H^{n}_{dR}(\bY)$.

It is clear that $\eta$ is globally variational if and only if $\del_{p} (\eta) = 0$. 
Every nonvanishing cohomology class in  $H^{p}_{dR}(\bY)$ gives rise to local variational problems:
two local variational problems of degree $p$ are {\em equivalent } if and only if they give rise to the same variational class of forms as the image of the corresponding morphism $\cE_p$ in the variational sequence. 

\section{Currents variationally associated with locally variational field equations}

We shall explicate now how cohomology enters in globality problems concerned with conserved quantities. We shall examine this aspect within Noether formalism \cite{Noe18} (see \cite{FrPaWiHist13} for an epistemological  discussion). 
Noether's Theorems already pointed out the cohomological content of the invariance of the Euler-Lagrange expressions and dealt with the study of the relationship between currents associated with symmetries of {\em global} Lagrangians and of corresponding Euler-Lagrange expressions; see, in particular, \cite{BeHa21}.

At the classical (infinitesimal) level, Noether Theorems can be formulated by means of the Lie derivate of a Lagrangian density and in fact consist of deriving (by an integration by parts procedure) a suitable version of the Cartan formula for the Lie derivative of horizontal $n$-forms on prolongations of fibered manifolds. We notice that the underling concept is that of finding a Lepage equivalent \cite{Lep36} for which a suitable Cartan formula can be written, then factorize modulo contact structures \cite{Kru08,KrKrSa10,KrSe05,olga09,MuLe09} (this procedure is involved with the Legendre transformation and thus with the appearing of a {\em momentum} at any degree of forms).
This is the basic idea underling the definition of  a 
{\em variational Lie derivative} operator $\cL_{j_{r}\Xi}$ and of a {\em quotient Cartan formula} (a {\em variation  formula}) defined  on the sheaves of the variational sequence. This enable us to define symmetries of classes of forms of any degree in the variational sequence and corresponding  conservation theorems; see \cite{FPV02} and \cite{PaDGA13} for details.

Having a suitable version of the Cartan formula and using naturality properties,
for $\eta\in\Var^{p}_{r}$ a locally variational $p$-form such that $\del_{p}(\eta_\lam) \neq 0$, we have that $\del_p (\cL_{\Xi}\eta_{\lam}) = 0$, \ie the variational Lie derivative `trivializes' cohomology classes \cite{FePaWi10,PaWi12,PaWiGa12}.
This suggest us that, if we know that the first variational derivative of a local presentation is closed, then the second variational derivative define a trivial cohomology class. 
In fact, the first variational derivative -- with respect to symmetries of Euler-Lagrange expressions --  of a local presentation is closed; therefore a trivial cohomology class is defined by the variational derivative of currents \cite{FrPaWi12,FrPaWi13}.

Let  then $\eta_{\lam}$ be a global Euler--Lagrange morphism for a local variational problem $\lam_i$; within the previously stated geometric framework, Noether's First Theorem reads
\beq
\cL_{\Xi} \lam_{i}= \Xi_V\rfloor\eta_\lam +d_H \eps_i \,.
\eeq
Analogously, a suitable representation of the Lie derivative of classes of dynamical forms can be obtained by 
a quotient Cartan formula; for details, see \cite{PaDGA13}. It is perhaps worthy of mention that this is related with an integration by parts formula for $\Xi_V\rfloor\eta_\lam$; such relationship has been worked out for gauge-natural theories, 
see \eg  \cite{FFPW08,FFPW11,FrPaWi05,PaWi03,PaWi04,PaWi07,PaWi12b}.

Since we assume $\eta_\lam$ to be closed, the quotient Cartan formula reduces to $\cL_{\Xi} \eta_\lam =\cE_n (\Xi_V\rfloor \eta_\lam)$, and  if $\Xi$ is such that $\cL_{\Xi} \eta_\lam =0$, then $\cE_n (\Xi_V\rfloor \eta_\lam) = 0$;
therefore {\em locally} 
\beq
\Xi_V\rfloor\eta_\lam =d_H\nu_i\,.
\eeq 
Notice that $\mathfrak{d}(\Xi_V\rfloor \eta_{\lam}) = 0$, but in general we have the obstruction 
\beq
\del^{'}(\Xi_V\rfloor \eta_{\lam}) \equiv \del_{n-1}(\Xi_V\rfloor \eta_{\lam})\neq 0\,,
\eeq 
so that the current  $\nu_i$ is a {\em local} object and it is conserved along the solutions of Euler--Lagrange equations (critical sections) \cite{FePaWi10}.

On the other hand, {\em and  independently} (see \cite{Noe18}), we get {\em locally}
\beq
\cL_{\Xi} \lam_{i}=d_H \beta_{i}\,,
\eeq
thus  we can write
$\Xi_{V} \rfloor \eta_{\lam}  + d_{H}( \eps_i  -  \beta_{i} )$  $=$ $0$, where $\eps_i$ is the usual {\em canonical} Noether current. 
\bDf
We call the (local) current $\eps_{i} - \beta_{i}$ a {\em Noether--Bessel-Hagen current}.
\eDf
Note that when $\Xi$ is only a symmetry of a dynamical form but not a symmetry of the Lagrangian, the so-called {\em strong Noether current} $\nu_i + \eps_i$ {\em is not a conserved current} and it is such that $d_H (\nu_i+\eps_i)$ is locally equal to $d_H\beta_i$ (\ie they are variationally equivalent).

At this point it deserves attention that we are studing conserved quantities associated with symmetries of global dynamical forms being global Euler-Lagrange forms of local presentations. 
We identified an obstruction to globality of such currents and it is a natural question whether  and when  we could instead be able to associate global conserved currents with globally invariant dynamics. It is natural to expect that somehow some restrictions have to be made. We can, for example, proceed in two different manner: 
\begin{enumerate}
\item look for an additional principle inspired by the fact that $\del^{'} (\cL_{\Xi} (\Xi_V\rfloor \eta_{\lam}))=0$ and therefore restrict the space of ``admissible'' generalized symmetries;

\item look for conditions on the cohomology of the configuration space and on the existence of global critical sections and therefore restrict the ``admissible'' fields.
\end{enumerate}

The first approach provided us with the following results \cite{FrPaWi12,FrPaWi13}.
\bDf
A conserved current {\em variationally associated with locally variational invariant field equations}  is a current associated with {\em a symmetry of the field equation which  is also required to be a symmetry for the variational derivative of  the local problem generating such a field equation}, \ie a current associated to a vector field $\Xi$ satisfying 
\beq
\cL_{\Xi}\eta_\lam = 0 \quad \textstyle{and} \quad   \cL_{\Xi}\cL_{\Xi}\lam_i = 0\, \quad \textstyle{but}  \quad  \cL_{\Xi}\lam_i\neq 0\,.
\eeq
\eDf
We see that the condition $\cL_\Xi \cL_\Xi \lam_i =0$ implies only $\mathfrak{d}d_H\nu_i=0$, \ie $d_H\nu_i$ is global. 

If we require the stronger condition $\cL_{\Xi}\mathfrak{d}\lam_i = 0$ (\ie the Lie derivative drags the local problems in such a way that their variational derivative coincide on the intersections of two open sets), we get $\mathfrak{d}d_H\bet_i=0$, \ie  the divergence of the strong Noether current is {\em global}.
This is also equivalent to say that the coboundary of the strong Noether currents is locally exact.
Under this condition  we have the conservation law $d_H \cL_\Xi  (\nu_i+\eps_i)=0$,  where $\cL_\Xi  (\nu_i+\eps_i)$, the {\em variation of the strong Noether currents},  is a {\em  local representative} of a global conserved  current.

\section{Cohomological obstructions and existence of global critical sections}

In this section we now investigate what is possible to state following the second approach.
We have seen that what we called the Noether--Bessel-Hagen current  $\eps_{i} - \beta_{i}$ is a {\em local} object and it is a current associated with a generalized symmetry (conserved along critical sections, since we locally have  $\nu_i = \eps_i - \beta_i + d_H \psi_{ij}$).  
In the following we shall denote by $[ \cdot , \cdot ]$ a class in de Rham cohomology. In the previous section, we saw that the current $\nu_i$ (as well as the Noether-Bessel-Hagen current) is variationally equivalent to a global (conserved) current if and only if $0 = [\Xi_{V} \rfloor \cE_{n}(\lam_{i})] \in H^{n}_{dR}(\bY)$.

We shall  investigate the relationship between the existence of global critical sections and the existence of global Noether--Bessel-Hagen currents. We  show that for a large class of theories, if a given local variational problem is supposed to admit global solutions, the above cohomology class is trivial. Vice versa, we obtain a new type of cohomological obstruction to the existence of global critical sections for a variational problem.

In the following we shall need to see certain cohomology classes in $H^{n}_{dR}(\bY)$ as the pull-back of cohomology classes in $H^{n}_{dR}(\bX)$, \ie defined by {\em closed differential forms on the base manifold}. For this reason we shall assume some restrictive conditions, which in fact can be relaxed; for details see \eg \cite{Win13}.
Note that $J_r\bY$ and $\bY$ have the same de Rham cohomology; accordingly, to simplify the notation, we shall omit the jet prolongation order.
We also stress that our reasoning works well in cohomology in degree $n$, but not greater.

Let us then assume $\sig$ be a {\em global section} of $\pi: \bY\to\bX$; for any global section we have of course $\sig^{*}\circ \pi^{*} =\textstyle{id}_{H^{n}_{dR}(\bX)}$.  Let  furthermore $H^{n}_{dR}(\bY) \sim  \pi^{*} (H^{n}_{dR}(\bX))$, \ie the pull-back $\pi^{*}: H^{n}_{dR}(\bX) \to H^{n}_{dR}(\bY)$ is an isomorphism of cohomology groups.

\bPr
Under the assumption of the isomorphism above, if a given (local) variational problem admits global critical sections (solutions),
all conservation laws derived from symmetries of global field equations admit global Noether--Bessel-Hagen conserved currents.
\ePr

\bPf
Let then $\sig$ be a global {\em critical} section; being a critical section the locally variational source form $\eta$ vanishes along its corresponding jet prolongation $j \sig$ and we have that also $\Xi_{V} \rfloor \eta$  vanishes along $j \sig$.
Since it is a global section we get that its pull-back to $\bX$ is identically zero, so that $[ j \sig^{*} \Xi_{V} \rfloor \eta] =0$ and, hence, since $\pi^{*}$ is an isomorphism, we have $[\Xi_{V} \rfloor \eta]  = 0$.  
The vanishing of  $\Xi_{V} \rfloor \eta$ along $j \sig$ implies the vanishing of the pull-back of a differential form representing the cohomology class $[ \Xi_{V} \rfloor \eta]$ and hence of the image of the cohomology class in $H^{n}_{dR}(\bX)$. This works well in spite of the fact that $\eta$, and thus also $\Xi_{V} \rfloor \eta$, is not a differential form, but a representative of an equivalence class of differential forms in the quotient $\For^{*}_{r}/\Thd^{*}_{r}$.  

This is apparently a little problem, because we need to compare the cohomology of the base manifold with the cohomology of the total space, and while $H^{n}_{dR}(\bY)\simeq H^{n}_{VS}(\bY)$, in principle we do not have a similar isomorphism for the cohomology groups $H^{n}_{dR}(\bX)$. This is overcome by the peculiar structure of quotient space in the variational sequence: for all sections 
$j \sig^{*} (\Thd^{*}_{r})=0$, we can correctly define $j \sig^{*} [\eta ]  \doteq [ j \sig^{*} \eta ]$; in fact, as explained before since $\Thd^{n}_{r}$ consists of contact $n$-forms the pull-back $\sig ^{*}$ factorizes over $\For^{n}_{r}/\Thd^{n}_{r}$. It is furthermore a well known fact that we can always  find a closed form $\alp \in \For^{n}_{r}$ which represents the cohomology class $[ \Xi_{V} \rfloor \eta]$ and which projects onto $\Xi_{V} \rfloor \eta \in \For^{n}_{r} / \Thd^{n}_{r}$. Thus,  $j \sigma ^{*} (\alpha) = 0$ and the corresponding class vanishes in $H^{n}_{dR}(\bX)$; therefore $[\Xi\rfloor \eta_\lam] = 0$, then there exist global conserved currents 
for all generalized symmetry $\Xi$.
 \ePf
Note that the obstruction does not depend on the section because of the isomorphism.

We notice that without the assumption of the isomorphism of cohomology groups, we can anyway  say that  $[\Xi\rfloor \eta_\lam] \notin \pi^{*} H^{n}_{dR} (\bX)$.  Of course, in this case we cannot infer in the same way the existence of global currents.

\bEx
In the case of the geometric formulation of  Chern-Simons theories $\bY$ takes the form $J^{1}\bP/\bG$ for a principal bundles $\bP$ over an arbitrary odd dimensional $\bX$.  Since $J^{1}\bP/\bG$ is an affine bundle over $\bX$, $J^{1}\bP/\bG$  and $\bX$ are homotopy equivalent and, in particular, $H^{n}_{dR} (J^{1}\bP/\bG) \sim H^{n}_{dR}(\bX)$. 
We can therefore state that Chern Simons theories, which admit global critical sections,  always admit global conserved quantities (Noether--Bessel-Hagen currents).
\eEx

We can use the  result above also conversely to define a cohomological obstruction to the existence of global solutions for a given problem. 

\bCr
Let $\sigma$ be a section of  $\bY$ over $\bX$. Let $\alp$ be a closed differential form which represents 
$[ \Xi_{V} \rfloor \eta]$, the obstruction  to the existence of global conserved quantities for conservation laws associated with the problem $\eta$ and its symmetry $\Xi$, in $H^{n}_{dR}(\bY)$. If the class  $j \sig^{*} [ \alp]\in H^{n}_{dR}(\bX)$ does not vanish, then there are no global solutions in the homotopy class of $\sig$.
\eCr

Now, as discussed before, if $\pi: \bY \to \bX$ admits a global section $\sig$, then this section (as well as any other section) defines a projection  $ \sig^{*}: H^{n}_{dR}(\bY)  \mapsto H^{n}_{dR}(\bX)$; the non vanishing of  $j \sig^{*}([ \Xi_{V} \rfloor \eta])$ is an obstruction to the existence of global solutions in the homotopy class of $\sig$. 
Note that, in this case, the obstruction {\em does depend} on the section; $j \sig^{*} [ \alp]$ may vanish for sections from one homotopy class, but not for those from another.

It is also important to note that  this obstruction is not identically zero, in general. 
We remark that  in the case when the isomorphism of cohomology groups is not given, if a global section is not critical, then  the class $[ \Xi_{V} \rfloor \eta ]$ is an obstruction for any other homotopic section to be a critical section \cite{Win13}.

\subsection*{Acknowledgement}

Research supported by Department of Mathematics, University of Torino. 
One of us (E. W.) has been partially supported by Department of Mathematics, University of Ostrava.


\end{document}